\documentclass[reprint,aps]{revtex4}
\usepackage{amsmath}
\usepackage{graphicx}
\usepackage{dcolumn}
\usepackage{bm}

\begin{document}

\title{Symmetric and antisymmetric nonlinear modes supported by dual local
gain in lossy lattices}
\author{K. W. Chow,$^{1}$ Edwin Ding,$^{2}$ Boris A. Malomed,$^{3}$ and A.
Y. S. Tang$^{1}$}

\address{ $^{1}$ Department of Mechanical Engineering, University of Hong
Kong, Pokfulam Road, Hong Kong\\
$^{2}$Department of Mathematics and Physics, Azusa Pacific
University, Box
7000, Azusa, CA 91702-7000, USA\\
$^3$ Department of Physical Electronics, School of Electrical
Engineering, Faculty of Engineering, Tel Aviv University, Tel Aviv
69978, Israel}

\begin{abstract}
We introduce a discrete lossy system, into which a double \textquotedblleft
hot spot" (HS) is inserted, i.e., two mutually symmetric sites carrying
linear gain and cubic nonlinearity. The system can be implemented as an
array of optical or plasmonic waveguides, with a pair of amplified nonlinear
cores embedded into it. We focus on the case of the self-defocusing
nonlinearity and cubic losses acting at the HSs. Symmetric localized modes
pinned to the double HS are constructed in an implicit analytical form,
which is done separately for the cases of odd and even numbers of
intermediate sites between the HSs. In the former case, some stationary
solutions feature a W-like shape, with a low peak at the central site, added
to tall peaks at the positions of the embedded HSs. The special case of two
adjacent HSs is considered too. Stability of the solution families against
small perturbations is investigated in a numerical form, which reveals
stable and unstable subfamilies. The instability generated by an isolated
positive eigenvalue leads to a spontaneous transformation into a co-existing
stable antisymmetric mode, while a pair of complex-conjugate eigenvalues
gives rise to persistent breathers.

\emph{This article is a contribution to the volume dedicated to Professor
Helmut Brand on the occasion of his 60th birhday}.
\end{abstract}

\maketitle


\section{Introduction}

Dissipative solitons, which originate, in the spatial domain, from the
simultaneous balance between diffraction and self-focusing, and between gain
and loss, are the subject of fundamental importance in optics~\cite{Rosanov}
and plasmonics~\cite{plasmonics,Marini}. An obvious condition necessary for
the stability of dissipative bright solitons is the stability of the zero
solution, i.e., background around the soliton. This condition may be
satisfied by linearly coupled complex Ginzburg-Landau (CGL) equations~\cite%
{wg1}, which describe dual-core waveguides, with the linear gain applied to
the active core, while the parallel-coupled passive one is lossy~\cite%
{dual,Pavel,Chaos,Marini}. The stability of dissipative solitons can be also
provided by the single CGL equation including the linear loss, cubic gain
and quintic loss, as shown in detail in many earlier works \cite{CQ,cqgle}
and some recent ones \cite{CQ-recent}.

Lately, considerable attention was drawn to another method of the creation
of stable localized dissipative modes, using linear gain localized at a
\textquotedblleft hot spot" (HS)~\cite{Mak,HSexact,HS,Valery}.
Configurations with multiple HSs~\cite{spotsExact1,spots,spotsExact2}, as
well as with extended amplifying structures~\cite{Zezyu}, have been
introduced too. Such settings can be created by implanting gain-producing
dopants into one or several narrow segments of the waveguide~\cite{Kip}, or
by focusing an external pumping beam at the target spot(s) in a
uniformly-doped waveguide.

Solutions for dissipative solitons pinned to narrow HSs approximated by
delta-functions have been found in an analytical form \cite%
{HSexact,spotsExact1,spotsExact2}. More sophisticated one- and
two-dimensional localized modes, such as vortices supported by the gain
acting in a ring-shaped area~\cite{2D}, have been found in the numerical
form~\cite{HS,Valery,spots}.

A natural setting for the implementation of the HS is provided by discrete
systems, i.e., lossy multi-core waveguiding arrays, where the gain is
applied to a single selected core. Assuming that the nonlinearity is also
concentrated at the pumped core, while the bulk of the array is linear,
exact solutions for modes pinned to the HS in such a system were recently
found in an implicit analytical form in Ref. \cite{we}, and stability
boundaries for the solutions were identified in a numerical form, in the
parameter plane of the linear gain and cubic loss (or gain) acting at the
HS. In particular, it was demonstrated that the pinned modes may be \emph{%
stable}, under the combined action of the \emph{unsaturated} cubic gain (in
the absence of quintic losses) and cubic self-defocusing nonlinearity, which
is impossible in uniform dissipative media (but possible in the case of the
localized unsaturated cubic gain acing in a continuous medium \cite{Valery}%
). On the other hand, it was demonstrated that the interplay of the linear
gain, cubic loss, and cubic self-defocusing nonlinearity in the same system
gives rise to a bistability, which is a noteworthy effect too.

The next natural step is to consider the discrete system with two mutually
symmetric HSs embedded into it, which is the subject of the present work.
Previously, symmetric modes pinned to a pair of symmetric nonlinear sites
implanted into a linear lattice were studied in Ref. \cite{embed}, and exact
solutions for stable asymmetric modes in the same system, generated by a
symmetry-breaking bifurcation, were recently reported in Ref. \cite{embed2}.

The paper is organized as follows. The discrete CGL equation with the
embedded double HSs, and its implicit analytical solutions for pinned modes
are introduced in Sec.~II. Numerical results, which include the
linear-stability analysis of the analytically found stationary solutions and
direct simulations of the underlying discrete CGL equation, are reported in
Sec. III. Because the system features the competition between the gain and
loss, direct simulations converge to attractors. In particular, the
simulations allow us to identify stable modes into which unstable
analytically found ones spontaneously transform. The paper is concluded by
Sec. IV.

\section{The model and analytical results}

\subsection{Formulation of the model}

We consider the transmission of optical or plasmonic waves in a discrete
array of linear lossy waveguides, with two identical nonlinear pumped cores
(alias HSs) embedded into it:
\begin{gather}
\frac{\mathrm{d}u_{m}}{\mathrm{d}z}=\frac{i}{2}\left(
u_{m-1}-2u_{m}+u_{m+1}\right) -\gamma u_{m}  \notag \\
+\left[ \left( \Gamma _{1}+i\Gamma _{2}\right) +\left( iB-E\right)
|u_{m}|^{2}\right] \left( \delta _{m,a}+\delta _{m,b}\right) u_{m}\;,
\label{eq:gl}
\end{gather}%
where\ $z$ is the propagation distance, $m$ the number of the waveguiding
core, $u_{m}(z)$ the complex amplitude of the electromagnetic field in it, $%
\gamma \geq 0$ the linear-loss parameter, and the coefficient of the linear
coupling between adjacent cores is scaled to be $1$. Further, $\Gamma _{1}>0$
is the linear gain applied at the two HSs, which are represented by cores
with numbers $a$ and $b$ (both numbers are integer), $\Gamma _{2}\geq 0$ is
the attractive potential which may be a part of the HS, $B>0$ or $B<0$
characterize the self-focusing or defocusing Kerr nonlinearity acting in the
active cores, and $E>0$ is the coefficient of the cubic loss. In light of
the analysis performed in Ref. \cite{we} for the discrete array with the
single embedded HS, we focus below on the self-defocusing nonlinearity,
fixing $B=-1$, as otherwise the pinned modes are prone to be unstable. We
present the analysis separately for the cases of odd and even numbers of
sites separating the two HSs.

Numerical simulations of Eqs. (\ref{eq:gl}) were performed in a sufficiently
large finite domain, whose size was greater than the width of the pinned
modes (for example, the computation domain was $|m|~\leq 30$ in the case
shown below in Fig. \ref{fig1}). In the simulations, the equations at the
edge sites (e.g., at $m=\pm 30$ in this example) were solved dropping the
fields corresponding to nonexisting sites (i.e., formally setting $u_{\pm
31}\equiv 0$ in the above-mentioned case).

\subsection{An odd number of intermediate sites between the hot spots}

To model the scenario with an odd number of sites between the two HSs, we
set in Eq. (\ref{eq:gl}) $a=N_{0}$ and $b=-N_{0}$, where $N_{0}=1,2,3,...$
is a positive integer, and the number of the intermediate sites is $2N_{0}-1$%
. In this case, we look for a stationary solution to Eq. (\ref{eq:gl}) in
the form of
\begin{equation}
u_{m}=U_{m}e^{ikz},  \label{U}
\end{equation}%
where a piecewise ansatz for the symmetric double-peak mode is
\begin{equation}
U_{m}=\left\{
\begin{array}{c}
Ae^{-\lambda |m|},\;\mathrm{if~~}|m|\geq N_{0}\;, \\
C\cosh (\lambda m),\;\mathrm{if}\;|m|<N_{0}\;.%
\end{array}%
\right.  \label{eq:ansatz1}
\end{equation}%
Here $\lambda \equiv \lambda _{r}+i\lambda _{i}$ is a complex eigenvalue,
with $\lambda _{r}>0$, and amplitude $A$ may be defined to be real.
Substituting this ansatz into Eq~.(\ref{eq:gl}) gives the following system
of algebraic equations:
\begin{gather}
\cosh \lambda _{r}\cos \lambda _{i}-1=k\;,  \notag \\
\sinh \lambda _{r}\sin \lambda _{i}=-\gamma \;,  \notag \\
Ae^{-\lambda N_{0}}=C\cosh (\lambda N_{0})\;,  \notag \\
\Gamma _{1}+i\Gamma _{2}+(iB-E)|A|^{2}e^{-2\lambda _{r}N_{0}}=\frac{i}{2}%
\left( \frac{e^{\lambda }-e^{-\lambda }}{1+e^{-2\lambda N_{0}}}\right) \;.
\label{eq:sym1}
\end{gather}%
This system, and its counterpart (\ref{eq:sym2}), derived below for the case
of an even distance between the HSs, are referred to as \textit{reduced
models} in the rest of the paper. Equations (\ref{eq:sym1}) and (\ref%
{eq:sym2}) were solved numerically by means of the Newton's method.

Note that, in addition to the symmetric \ modes represented by ansatz (\ref%
{eq:ansatz1}), it is also possible to look for antisymmetric solutions, in
the form of
\begin{equation}
U_{m}=\left\{
\begin{array}{c}
A~\mathrm{sgn}(m)~e^{-\lambda |m|},\;\mathrm{if~~}|m|\geq N_{0}\;, \\
C\sinh (\lambda m),\;\mathrm{if}\;|m|<N_{0}\;.%
\end{array}%
\right.   \label{anti}
\end{equation}%
An obvious difference between \textit{ans\"{a}tze} (\ref{eq:ansatz1}) and (%
\ref{anti}) is that the former one always has a nonzero amplitude at the
center, $U_{0}\equiv C\neq 0$, while the antisymmetric solution
automatically vanishes at the central site ($U_{0}=0$). Antisymmetric modes
will be considered in detail elsewhere, although numerically found examples
are presented below. They emerge as a result of the spontaneous
transformation of unstable symmetric modes, see the profiles depicted by the
blue continuous line in the top left top panel of Fig. \ref{fig4}, and
middle left panel of Fig. \ref{fig8}.

\subsection{An even number of intermediate sites between the hot spots}

In the case of an even number of sites between the HSs, we set in Eq. (\ref%
{eq:gl})
\begin{equation}
a=N_{0},~b=-N_{0}+1,  \label{N}
\end{equation}%
where $N_{0}=1,2,3,$ ... is again a positive integer, the corresponding
number of intermediate sites between the HSs being $2\left( N_{0}-1\right) $%
. In this case we look for stationary solution (\ref{U}) by means of the
following piecewise ansatz, which are symmetric about the off-site central
point, $m=1/2$:

\begin{equation}
U_{m}=\left\{
\begin{array}{c}
A~e^{-\lambda \left( m-1/2\right) },\;\mathrm{if}\;m\geq N_{0}, \\
A~e^{\lambda \left( m-1/2\right) },\;\mathrm{if}\;\;m\leq -N_{0}+1\;, \\
C\cosh \left( \lambda \left( m-1/2\right) \right) ,\;\mathrm{if}%
\;-N_{0}+1<m<N_{0}\;.%
\end{array}%
\right.  \label{eq:ansatz2}
\end{equation}%
With regard to this condition, substituting~ansatz (\ref{eq:ansatz2}) into
Eq.~(\ref{eq:gl}) leads to the following reduced system of equations:
\begin{gather}
\cosh \lambda _{r}\cos \lambda _{i}-1=k\;,  \notag \\
\sinh \lambda _{r}\sin \lambda _{i}=-\gamma \;,  \notag \\
A~e^{-\lambda (N_{0}-1/2)}=C\cosh \left( \lambda \left( N_{0}-\frac{1}{2}%
\right) \right) \;,  \notag \\
\Gamma _{1}+i\Gamma _{2}+(iB-E)|A|^{2}e^{-2\lambda _{r}(N_{0}-1/2)}=\frac{i}{%
2}\left( \frac{e^{\lambda }-e^{-\lambda }}{1+e^{-2\lambda (N_{0}-1/2)}}%
\right) \;.  \label{eq:sym2}
\end{gather}%
The first two equations in Eqs.~(\ref{eq:sym1}) are identical to those in
Eqs.~(\ref{eq:sym2}), as they are derived in the bulk lattice, off the HS
sites.

\section{The linear-stability analysis}

\label{sec:stab}

The stability of the pinned modes was studied by means of the standard
linearization procedure. To this end, perturbed solutions were taken as
\begin{equation}
u_{m}=\left[ U_{m}+\epsilon V_{m}(z)\right] e^{ikz},  \label{eq:ansatz}
\end{equation}%
where $V_{m}(z)\equiv X_{m}(z)+iY_{m}(z)$ is a complex perturbation with an
infinitesimal amplitude $\epsilon \ll 1$. Substituting this into Eq.~(\ref%
{eq:gl}), we derive the following linearized equations:%
\begin{eqnarray}
&&\frac{\mathrm{d}X_{m}}{\mathrm{d}z}=-\frac{1}{2}Y_{m-1}+(k+1)Y_{m}-\frac{1%
}{2}Y_{m+1}-\gamma X_{m}  \notag \\
&&+(\delta _{m,a}+\delta _{m,b})\left\{ \left( \Gamma _{1}X_{m}-\Gamma
_{2}Y_{m}\right) \right.  \notag \\
&&-B\left[ 2P_{m}Q_{m}X_{m}+\left( P_{m}^{2}+3Q_{m}^{2}\right) Y_{m}\right]
\notag \\
&&\left. -E\left[ \left( 3P_{m}^{2}+Q_{m}^{2}\right) X_{m}+2P_{m}Q_{m}Y_{m}%
\right] \right\} \;,  \notag \\
&&\frac{\mathrm{d}Y_{m}}{\mathrm{d}z}=\frac{1}{2}X_{m-1}-(k+1)X_{m}+\frac{1}{%
2}X_{m+1}-\gamma Y_{m}  \notag \\
&&+(\delta _{m,a}+\delta _{m,b})\left\{ \left( \Gamma _{2}X_{m}+\Gamma
_{1}Y_{m}\right) \right.  \notag \\
&&+B\left[ \left( 3P_{m}^{2}+Q_{m}^{2}\right) X_{m}+2P_{m}Q_{m}Y_{m}\right]
\notag \\
&&\left. -E\left[ 2P_{m}Q_{m}X_{m}+\left( P_{m}^{2}+3Q_{m}^{2}\right) Y_{m}%
\right] \right\} \;,  \label{eq:linear}
\end{eqnarray}%
where $P_{m}\equiv \mathrm{Re}(U_{m})$ and $Q_{m}\equiv \mathrm{Im}(U_{m})$.
An eigenvalue problem is obtained from here by substituting $X_{m}=\phi
_{m}\exp (\rho z)$ and $Y_{m}=\psi _{m}\exp (\rho z)$ into Eqs.~(\ref%
{eq:linear}), where $\rho $ is the instability growth rate. The pinned mode
is linearly stable provided that all the eigenvalues have $\mathrm{Re}\left(
\rho \right) \leq 0$.

\section{Numerical results}

\subsection{An odd number of intermediate sites between the hot spots}

\begin{figure}[tbph]
\begin{center}
\includegraphics[width = 80mm, keepaspectratio]{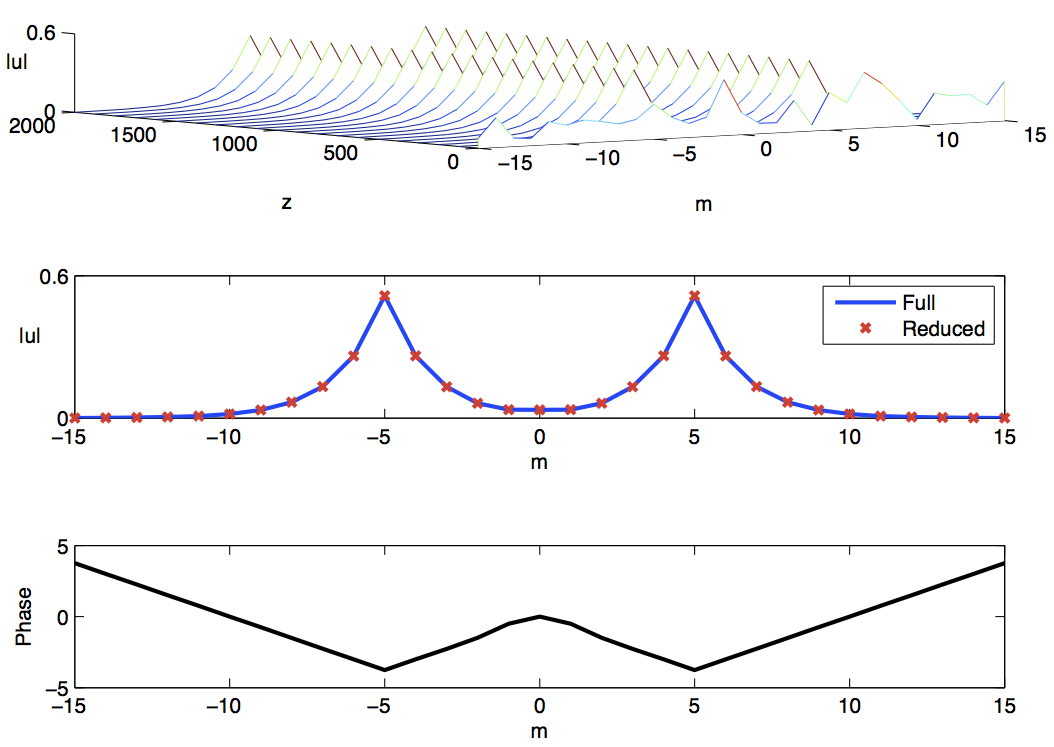}
\end{center}
\caption{(Color online) Top: Evolution of a random initial configuration
into a stable two-peak mode (\textit{attractor}) in the system with an odd
number of intermediate sites between the hot spots. The two hot spots are
located at $m=\pm 5,$ with parameters $B=-1$, $E=0.16$, $\Gamma _{1}=0.889$,
$\Gamma _{2}=0.8$, and $\protect\gamma =0.5$. The computational domain used
here is $-30\leq m\leq +30$. Middle: Absolute values of the solution
produced, as\ an \textit{attractor}, by direct simulations of Eq. (\protect
\ref{eq:gl}) (the blue solid line), and its counterpart obtained from the
reduced model~(\protect\ref{eq:sym1}) (red crosses). The blue curves and
chains of red crosses have the same meaning in similar figures displayed
below. Bottom: the phase structure of the numerically generated solution
(the reduced model produces the same phase profile).}
\label{fig1}
\end{figure}

The top panel of Fig.~\ref{fig1} shows a typical evolution of a discrete
pulse in the self-defocusing case, $B=-1$, with the cubic loss, $E>0$. When
linear gain $\Gamma _{1}$ exceeds a certain threshold value, a randomly
built initial profile evolves into a stable two-peak solution, with peaks
located at the HSs. The middle panel of Fig.~\ref{fig1} shows agreement
between the absolute values of the solutions produced by the direct
simulations (the blue solid line) and the results obtained from the reduced
system (\ref{eq:sym1}) (a chain of red crosses). In principle, one may
expect some discrepancy between the two profiles due to the fact that the
numerical solutions are obtained in a finite domain, but no real discrepancy
can be spotted in this and similar plots. The phase profile of the same
solution, whose gradient determines the energy flow in the established
state, is displayed in the bottom panel of Fig. \ref{fig1}.

\begin{figure}[tbph]
\begin{center}
\includegraphics[width = 80mm, keepaspectratio]{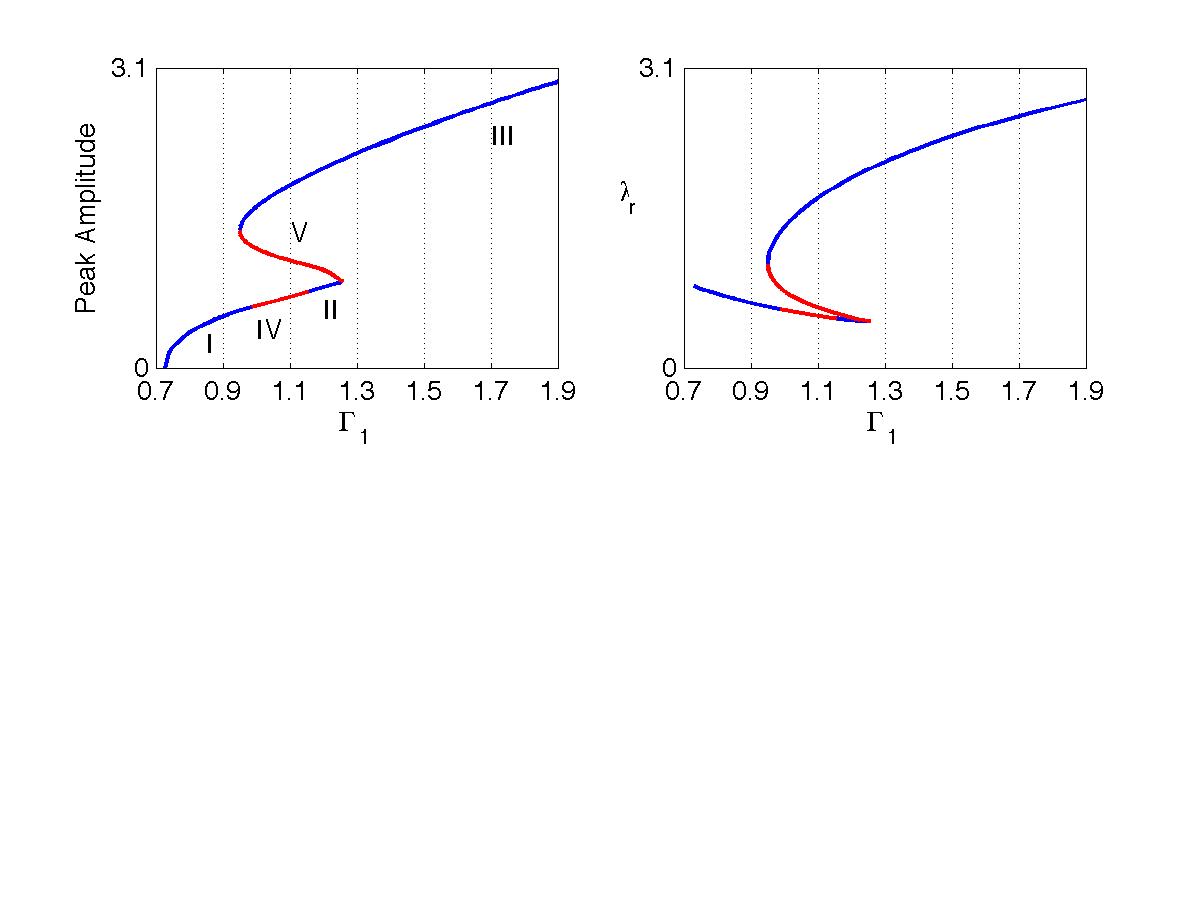}
\end{center}
\caption{(Color online) Characteristics of the family of symmetric
double-peak stationary solutions, obtained from the reduced system (\protect
\ref{eq:sym1}) in the model with the odd number of intermediate sites
between the hot spots, versus the linear gain, $\Gamma _{1}$. The parameters
are $B=-1$, $E=0.16$, $\Gamma _{2}=0.8$, $\protect\gamma =0.5$, and $N_{0}=5$%
. Here and in similar figures displayed below, blue and red segments
designate linearly stable and unstable solutions, respectively, as
identified from a numerical solution of the eigenvalue problem based on Eqs.
(\protect\ref{eq:linear}).}
\label{fig2}
\end{figure}

Next, we study the linear stability of the families of stationary pinned
modes. First, we build a branch of the symmetric double-peak solutions in
the form~of ansatz (\ref{eq:ansatz1}) by means of a continuation algorithm
applied to the reduced system (\ref{eq:sym1}). Figure~\ref{fig2} shows two
main characteristics of the solution family, namely the peak amplitude, $%
A\exp \left( -\lambda _{r}N_{0}\right) $ (see Eqs. (\ref{eq:ansatz1})), and
the inverse width, $\lambda _{r}$, as functions of the linear gain, $\Gamma
_{1}$. At $\Gamma _{1}\leq 0.7284$, which is the above-mentioned threshold
value, only the zero solution, with $A=C=0$, is possible, as the system does
not have enough gain to compensate the background loss and sustain any
nontrivial solution. Double-peak modes exist at $\Gamma _{1}>0.7284$. We
then computed the linear stability spectrum for each solution belonging to
the branch, which allows one to identify stable (blue) and unstable (red)
segments in Fig. \ref{fig2}. For the present set of the parameters, stable
solutions were found in the regions of $0.7284<\Gamma _{1}\leq 0.9863$
(region I), $1.157\leq \Gamma _{1}\leq 1.254$ (region II), and $\Gamma
_{1}\geq 0.9508$ (region III). Unstable branches were found in the
complementary segments, namely, $0.9863<\Gamma _{1}<1.157$ (region IV) and $%
0.9508<\Gamma _{1}<1.254$ (region V).

\begin{figure}[tbph]
\begin{center}
\includegraphics[width = 100mm, keepaspectratio]{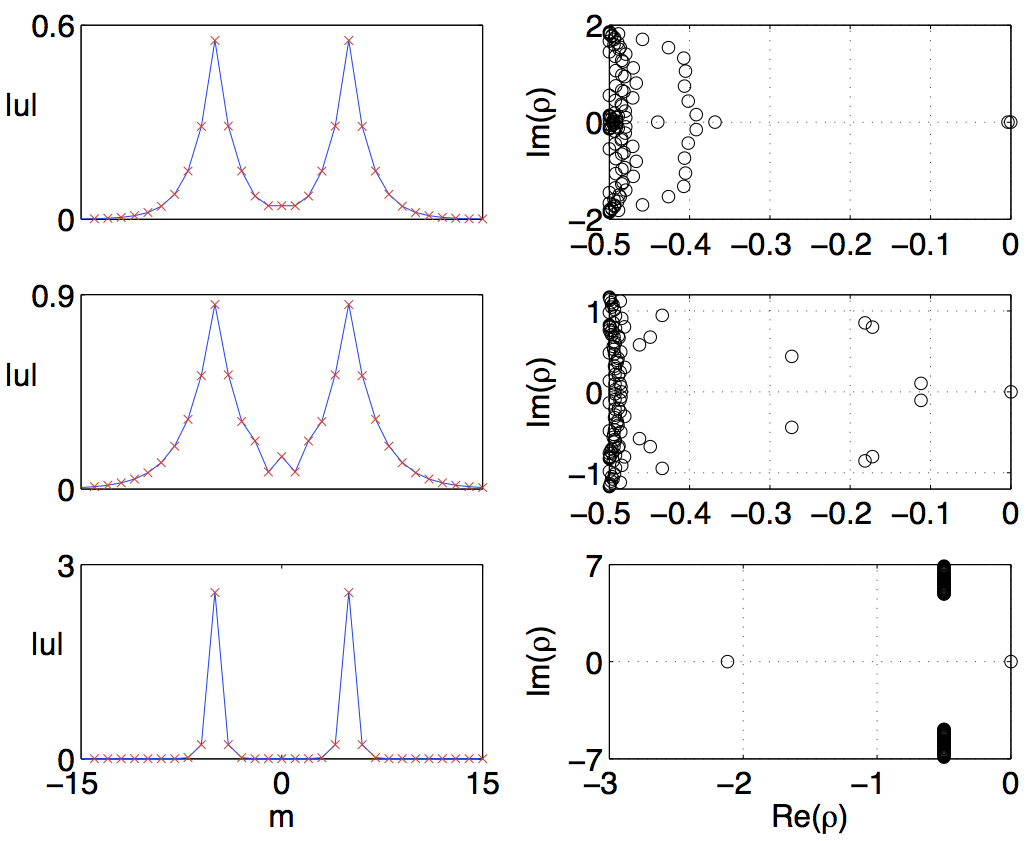}
\end{center}
\caption{(Color online) Left: Stable solutions produced by direct
simulations of Eq. (\protect\ref{eq:gl}), and by the reduced model (\protect
\ref{eq:sym1}), at $\Gamma _{1}=0.9179$ (top), $\Gamma _{1}=1.2246$
(middle), and $\Gamma _{1}=1.5627$ (bottom), in the model with an odd number
of intermediate sites. Other parameters are as in Fig. \protect\ref{fig2}.
Right: The corresponding linear-stability spectra.}
\label{fig3}
\end{figure}

Further, left plots in Fig.~\ref{fig3} display examples of solutions
obtained from direct simulations of Eq. (\ref{eq:gl}) (blue solid curves),
and their counterparts generated by reduced system (\ref{eq:sym1}) (red
crosses), in stable regions I, II, and III, while right plots display the
corresponding linear spectra. In stable region I (the top panel), the mode
features the simple profile, with the peaks existing solely at the HS
positions, $m=\pm 5$, and a minimum at the center ($m=0$). In contrast to
that, in stable region II (the middle panel), the steady-state solution
features a W-shaped profile, with an additional local peak appearing at $m=0$%
. Note that reduced system~(\ref{eq:sym1}) correctly predicts this more
sophisticated profile. In stable region III (the bottom panel), two tall
narrow peaks are, as a matter of fact, the same isolated solutions as those
recently found in the single-HS model \cite{we}, with negligible interaction
between them.

\begin{figure}[tbph]
\begin{center}
\includegraphics[width = 100mm, keepaspectratio]{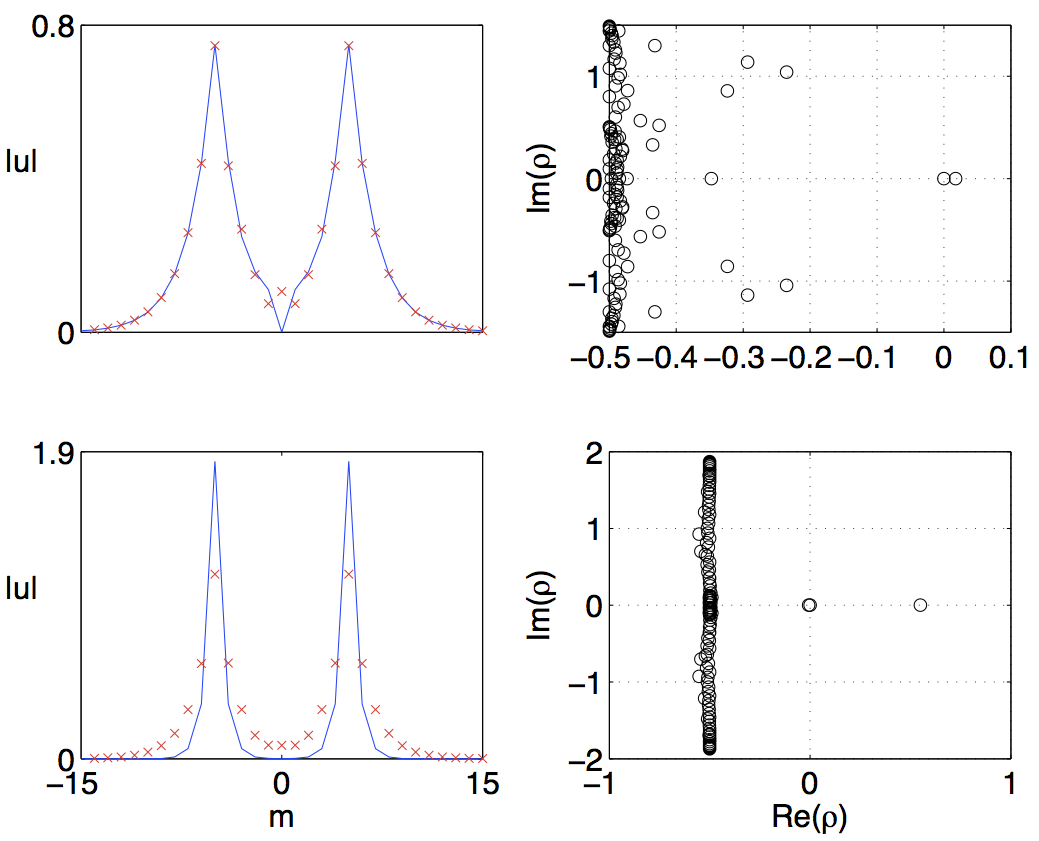}
\end{center}
\caption{(Color online) Left: The absolute value of unstable symmetric
solutions predicted by the reduced model with an odd number of intermediate
sites (chains of red crosses), and of the corresponding stable
(antisymmetric) solutions produced by direct simulations of Eq. (\protect\ref%
{eq:gl}) (blue solid curves) at $\Gamma _{1}=1.1186$ (top), and $\Gamma
_{1}=1.0724$ (bottom). Right: The corresponding linear-instability spectra
of the solutions obtained from the reduced system. Isolated eigenvalues with
$\protect\lambda _{i}=0$ and $\protect\lambda _{r}>0$ indicate
instabilities. }
\label{fig4}
\end{figure}

Figure~\ref{fig4} shows solutions of the reduced model (red crosses) which
turn out to be unstable, in segments IV and V of the solution branch (see
Fig.~\ref{fig2}). In the unstable region IV (the top row in the figure), the
solutions of the reduced system (red crosses) feature a W-shaped profile,
resembling the above-mentioned stable solution found in region II. However,
as these solutions are unstable, they cannot be obtained by direct
simulations of Eq. (\ref{eq:gl}). In fact, at the same parameters, the full
system evolves into a stable profile for which the amplitude at the central
site, $m=0$, is zero. Actually, this is a example of a stable\emph{\
antisymmetric} solution corresponding to ansatz (\ref{anti}), which will be
considered in more detail, including its analytical form, elsewhere. In
unstable region V (the bottom panels in Fig. \ref{fig4}), the two peaks in
the unstable solution predicted by the reduced system interact with each
other, featuring conspicuous overlap, whereas in the stable solution
generated by the direct simulations for the same parameter set, the peaks
are completely separated. In fact, in this case the stable solutions
generated by the direct simulations evolve into the modes which exist, at
the same parameters, in segment III on the stable branch (see Fig.~\ref{fig2}%
).

\begin{figure}[tbph]
\begin{center}
\includegraphics[width = 90mm, keepaspectratio]{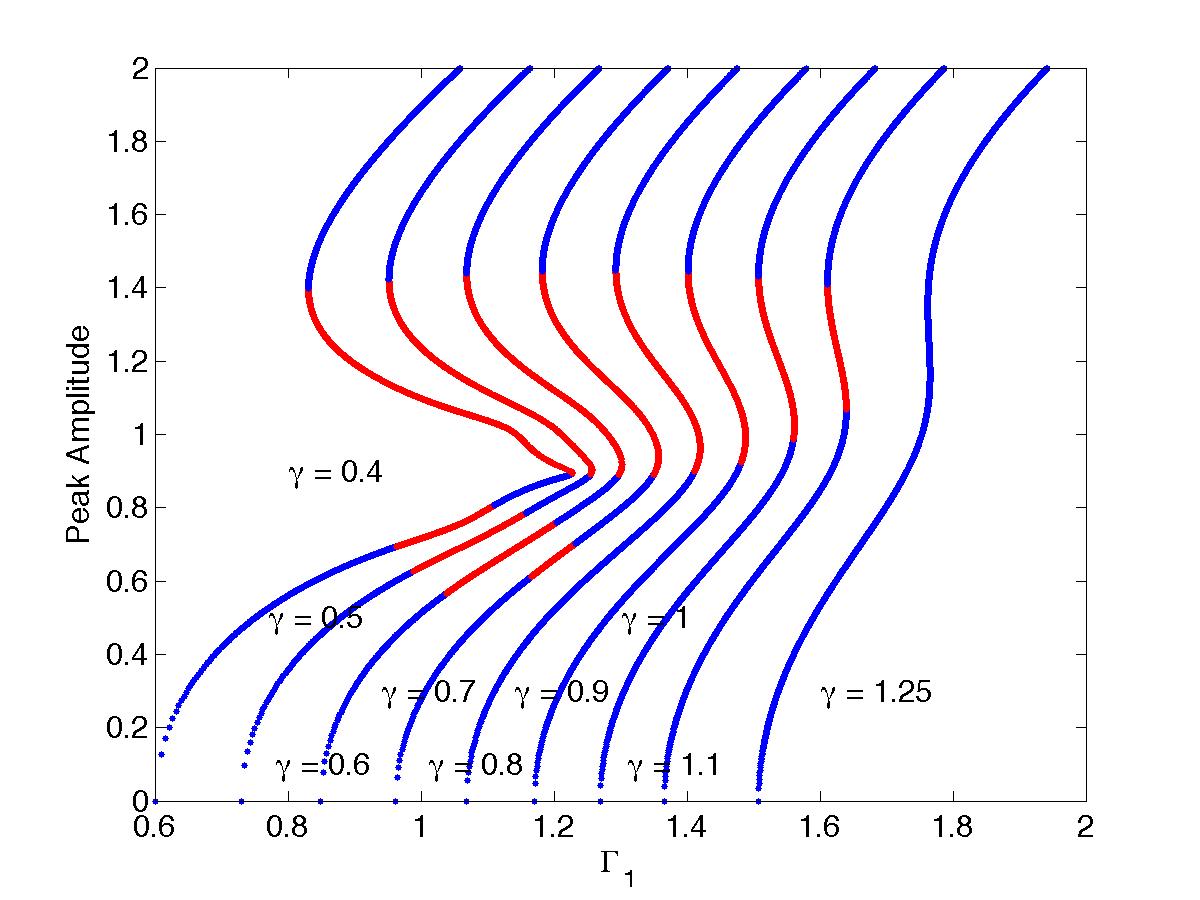}
\end{center}
\caption{(Color online) A set of solution branches found in the model with
the odd number of intermediate sites, at the following values of the
background-loss coefficient (from left to right, as indicated in the
figure): $\protect\gamma =0.4$, $0.5$, $0.6$, $0.7$, $0.8$, $0.9$, $1$, $1.1$%
, $1.25$. The parameters are $B=-1$, $E=0.16$, $\Gamma _{2}=0.8$, and $%
N_{0}=5$.}
\label{fig5}
\end{figure}

We also studied the linear stability for different values of the
linear-background loss coefficient, $\gamma $. Figure~\ref{fig5} shows a set
of solution branches for $\gamma $ varying from $0.45$ to $1.25$. At lower
values of the background loss (e.g., $0.4\leq \gamma \leq 0.7$) the branches
resemble the one shown above in Fig.~\ref{fig2}, i.e., they consist of three
stable and two unstable segments. When $\gamma $ increases, the unstable
segments shrink and eventually disappear. When this happens, the solutions
become stable for all values of $\Gamma _{1}$. Note that the fully stable
branches are single-valued, in contrast to multi-valued ones which include
unstable segments. In fact, this transition is a typical example of the
codimension-2 \textit{cusp bifurcation} \cite{cusp}

\begin{figure}[tbph]
\begin{center}
\includegraphics[width = 100mm, keepaspectratio]{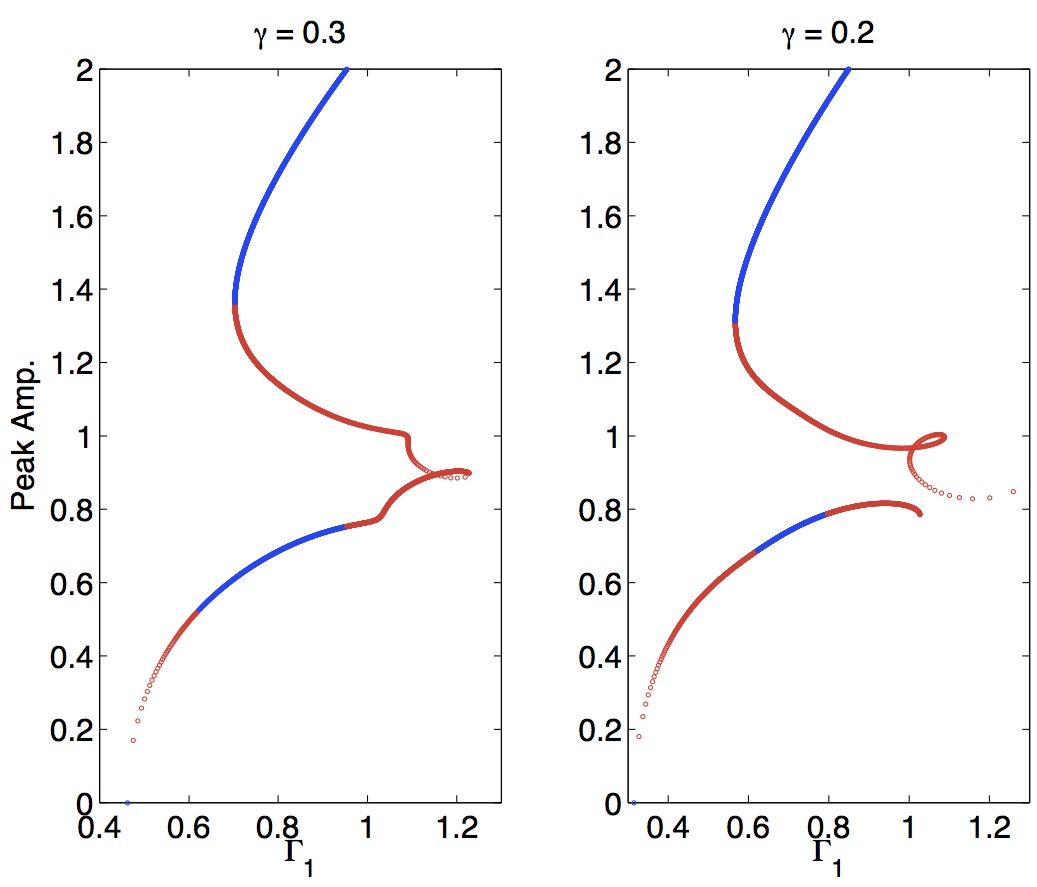}
\end{center}
\caption{(Color online) Solution branches at $\protect\gamma =0.3$ (left)
and $\protect\gamma =0.2$ (right) in the model with the odd number of
intermediate sites. Other parameters are as in Fig.~\protect\ref{fig5}.}
\label{fig6}
\end{figure}

\begin{figure}[tbph]
\begin{center}
\includegraphics[width = 90mm, keepaspectratio]{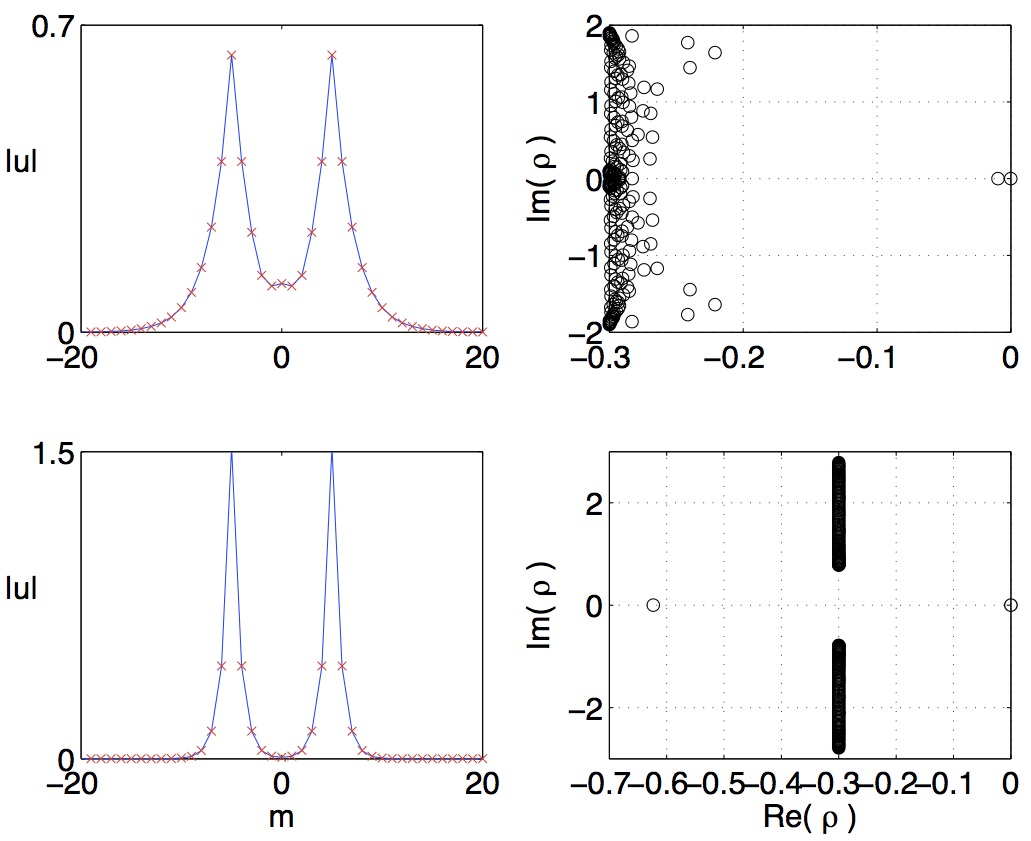}
\end{center}
\caption{(Color online) Left: Stable solutions in the model with the odd
number of intermediate sites at $\Gamma _{1}=0.7257$ and $\protect\gamma =0.3
$. Right: The corresponding linear-stability spectra.}
\label{fig7}
\end{figure}

\begin{figure}[tbph]
\begin{center}
\includegraphics[width = 90mm, keepaspectratio]{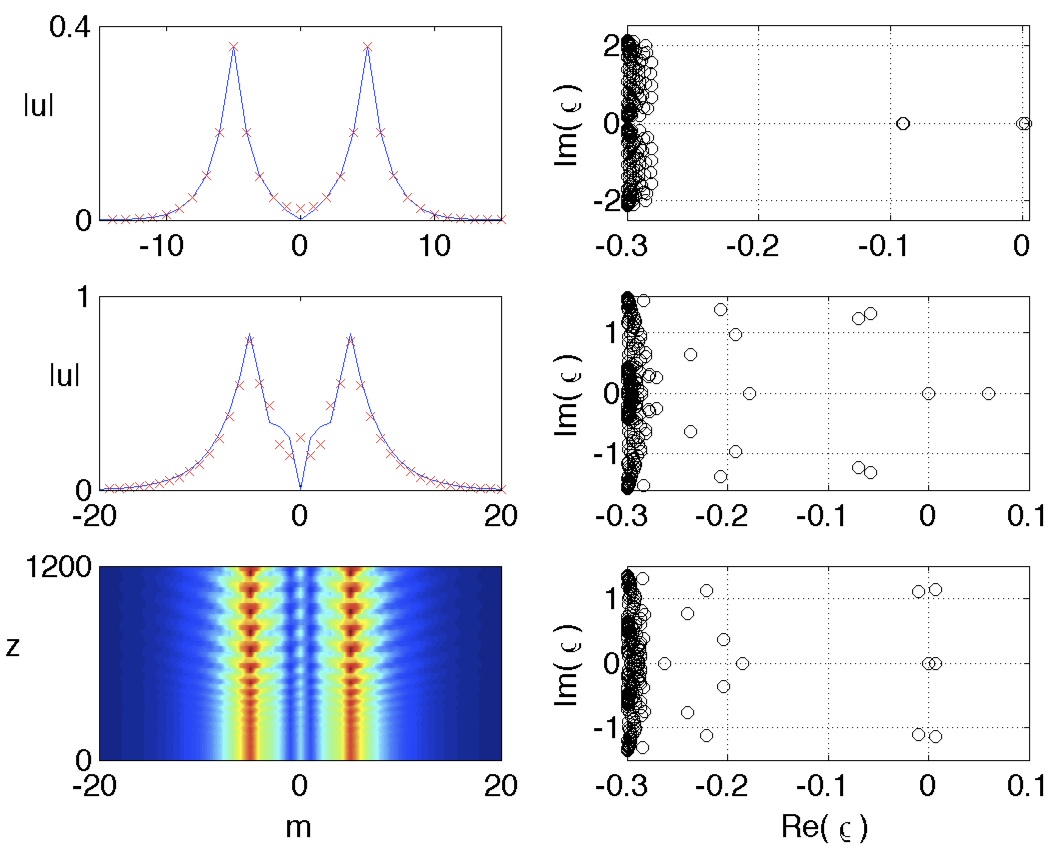}
\end{center}
\caption{Left: Unstable symmetric solutions produced by the reduced system,
and actual stable solutions generated by direct simulations of Eq. (\protect
\ref{eq:gl}) at $\protect\gamma =0.3$ and $\Gamma _{1}=0.5262$ (top), $%
\Gamma _{1}=1.0053$ (middle), and $\Gamma _{1}=1.0546$ (bottom). In the
latter case, the oscillatory instability transforms the stationary pinned
mode into a double-peaked breather. Right: The corresponding
linear-instability spectra for the solutions produced by the reduced system.}
\label{fig8}
\end{figure}

Figure~\ref{fig6} shows typical solution branches obtained for still smaller
levels of the background loss, $\gamma =0.3$ and $\gamma =0.2$ in the left
and right panels, respectively. For the present values of the parameters, we
have found two stable segments on the branches, with smaller and larger
amplitudes, severally. Two typical solution profiles and their
linear-stability spectra are shown in Fig.~\ref{fig7}, while Fig.~\ref{fig8}
shows unstable solutions belonging to the low-amplitude branch at $\gamma
=0.3$. The type of the instability revealed by the top and middle panels,
and represented by an isolated purely real positive eigenvalue, has already
been presented in Fig.~\ref{fig4}. However, a Hopf bifurcation, accounted
for by pairs of complex eigenvalues, is observed at larger $\Gamma _{1}$ on
the low-amplitude branch. Specifically, a pair of complex-conjugate
eigenvalues crosses the imaginary axis into the right (unstable) half-plane.
The Hopf bifurcation naturally leads to periodic oscillations, i.e.,
transformation of the unstable stationary mode into a persistent breather,
which keeps the overall double-peak structure, see an example in the bottom
row of Fig.~\ref{fig8}.

\begin{figure}[tbph]
\begin{center}
\includegraphics[width = 90mm, keepaspectratio]{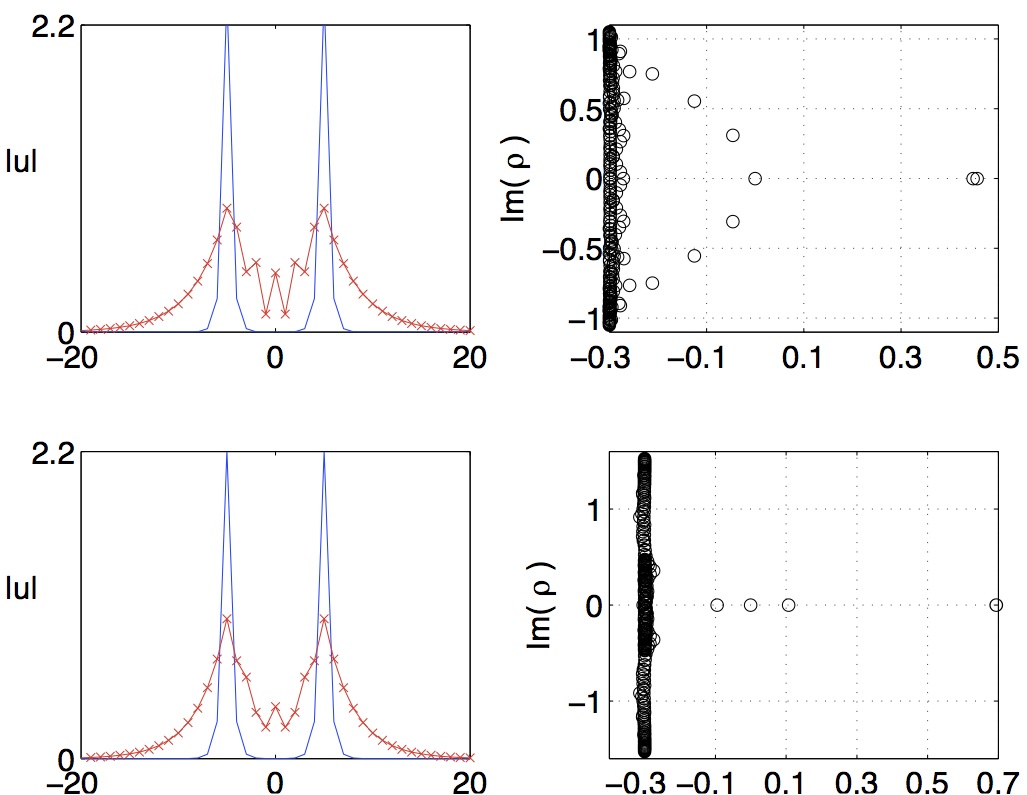}
\end{center}
\caption{(Color online) Left: Unstable solutions produced by the reduced
system in the model with the odd number of intermediate sites, and the
actual stable solutions obtained from direct simulations of Eq. (\protect\ref%
{eq:gl}) at $\protect\gamma =0.3$ and $\Gamma _{1}=1.219$ (top), and $\Gamma
_{1}=1.087$ (bottom). Right: The corresponding linear-instability spectra.}
\label{fig9}
\end{figure}

Note that the stable mode, into which the unstable one evolves in the middle
row of Fig. \ref{fig8}, is qualitatively similar to the one displayed in the
upper row of Fig. \ref{fig4}, also being an antisymmetric mode corresponding
to ansatz (\ref{anti}). Further, the evolution of unstable solutions
belonging to the high-amplitude branch leads to the establishment of two
fully isolated narrow peaks, see examples in Fig.~\ref{fig9}, which is
similar to the outcome of the instability development observed in the bottom
row of Fig.~\ref{fig4}. The instability of these solutions is accounted for
by an isolated positive eigenvalues in the linear spectrum, and the emerging
stable modes, which feature the isolated peaks, belong to the stable
high-amplitude segment, which exists at the same parameter values.

\begin{figure}[tbph]
\begin{center}
\includegraphics[width = 90mm, keepaspectratio]{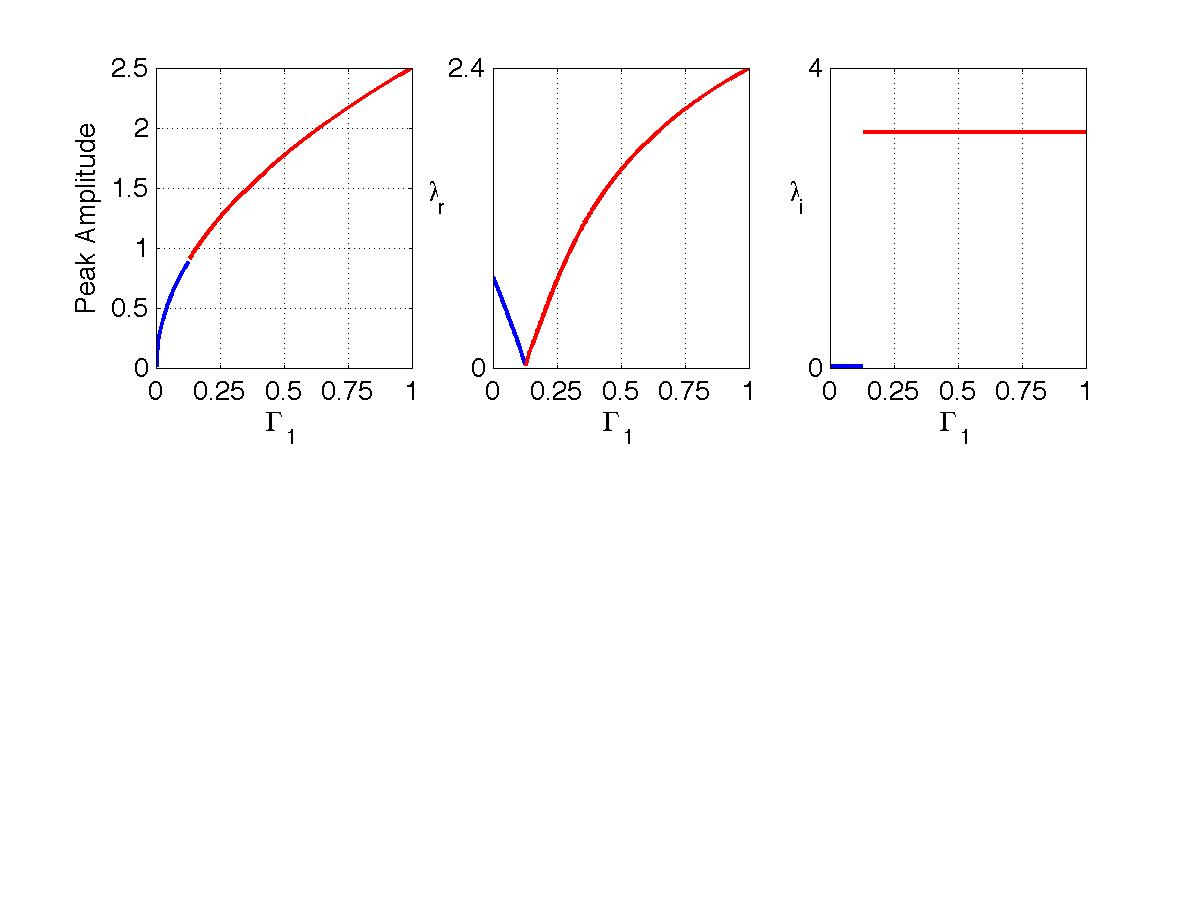}
\end{center}
\caption{(Color online) Characteristics of the solution branch in the model
with an odd number of intermediate sites and $\protect\gamma =0$ (no
background loss).}
\label{fig10}
\end{figure}

\begin{figure}[tbph]
\begin{center}
\includegraphics[width = 90mm, keepaspectratio]{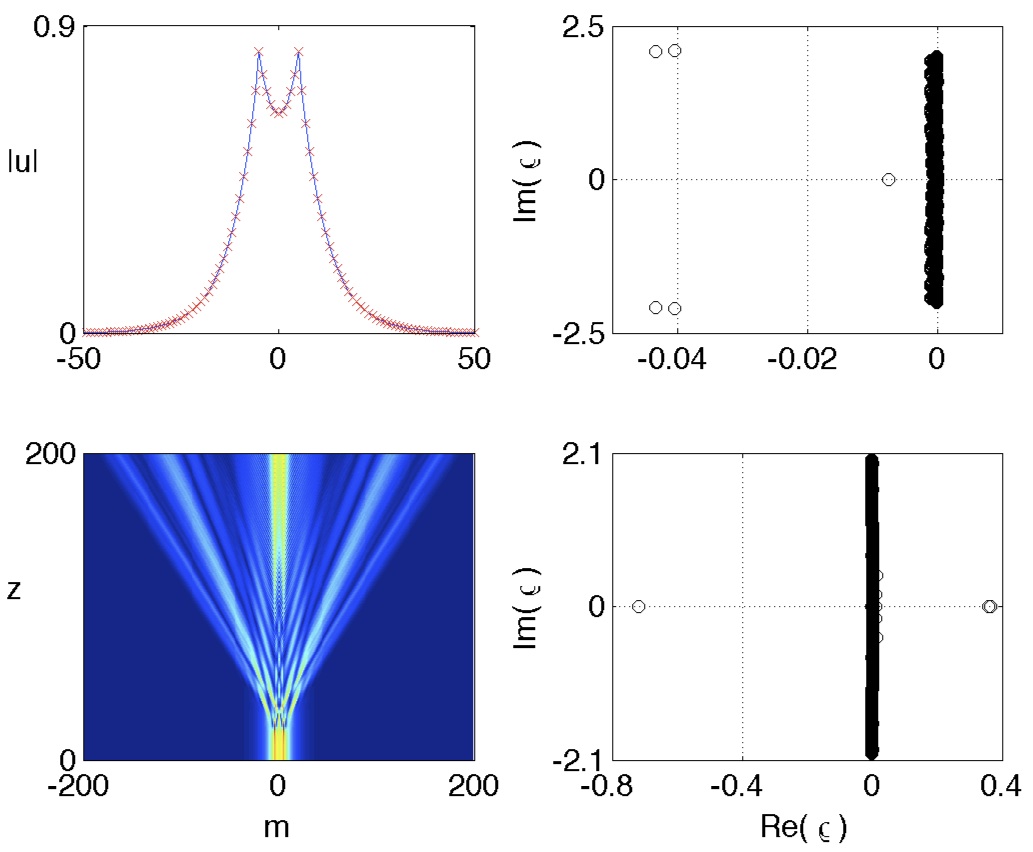}
\end{center}
\caption{(Color online) Left: Stable (top) and unstable (bottom) solutions
obtained from the full equation (\protect\ref{eq:gl}) and reduced model (%
\protect\ref{eq:sym1}) in the system with the odd number of intermediate
sites and $\protect\gamma =0$, for $\Gamma _{1}=0.1088$ and $\Gamma
_{1}=0.1434$, respectively. Right: The corresponding linear-stability
spectra.}
\label{fig11}
\end{figure}

When the background loss is absent ($\gamma =0$), solutions in the form
of~ansatz (\ref{eq:ansatz1}) can only be found in a small region of the
parameter space. Figure~\ref{fig10} shows a solution branch in this case.
Stable modes exist solely in the region of $0\leq \Gamma _{1}\leq 0.126$.
Note that, as it follows from the second equation in system (\ref{eq:sym1}),
the corresponding wavenumber $\lambda _{i}$ in ansatz (\ref{eq:ansatz1}) is
zero for stable solutions, and $\pi $ for unstable ones, i.e., the modes
with \textit{staggered} tails, corresponding to $\lambda _{i}=\pi $, are
unstable (in the numerical solution, $\lambda _{i}$ can converge to $n\pi $
with other integer values of $n$, but they all are tantamount to $n=0$ or $1$%
). The top row of Fig.~(\ref{fig11}) shows an example of a stable solution
obtained at $\Gamma _{1}=0.1088$, along with its linear spectrum. On the
other hand, the bottom row illustrates what happens in the unstable region.
In the absence of the background loss, secondary structures are generated in
the course of the evolution as a result of the emission of lattice waves
(\textquotedblleft phonons"). The positive eigenvalue in the spectrum is a
signature of the instability.

It is worth mentioning that, since $\lambda _{i}=0$ or $\lambda _{i}=\pi $
when $\gamma =0$, the real and imaginary parts of the last equation
in~system (\ref{eq:sym1}) give
\begin{eqnarray}
\Gamma _{1}/E &=&|A|^{2}e^{-2\lambda _{r}N_{0}}\;, \\
\Gamma _{2}+B|A|^{2}e^{-2\lambda _{r}N_{0}} &=&\pm \frac{\sinh \left(
\lambda _{r}\right) }{1+e^{-2\lambda _{r}N_{0}})}\;,
\end{eqnarray}%
where the plus and minus signs correspond to $\lambda _{i}=0$ and $\lambda
_{i}=\pi $, respectively. These equations can be combined to yield
\begin{equation}
\Gamma _{2}+\frac{B\Gamma _{1}}{E}=\frac{\pm \sinh \lambda _{r}}{%
1+e^{-2\lambda _{r}N_{0}}}\;.  \label{eq:exact1}
\end{equation}%
One can solve Eq. (\ref{eq:exact1}) for linear gain $\Gamma _{1}$ in terms
of $\lambda _{r}$, and, subsequently, the remaining solution parameters. $A$
and $C$, can be easily found from~the other equations of system (\ref%
{eq:sym1}).

\subsection{An even number of intermediate sites between the hot spots}

We have also studied the linear stability of solutions obtained in the form
of ansatz (\ref{eq:ansatz2}), whose parameters were found from a numerical
solution of reduced system~(\ref{eq:sym2}). For the same parameters as
adopted above (i.e., $N_{0}=5$, $B=-1$, $E=0.16$, $\Gamma _{2}=0.8$) and in
the presence of the background loss ($\gamma >0$), it has been found that
the solution branches and their stability resemble those reported in the
previous section (which is quite natural, as the two-peak modes of the two
types should be close for the numbers of the intermediate sites $N_{\mathrm{%
odd}}\equiv 2N_{0}-1=9$ and $N_{\mathrm{even}}\equiv 2\left( N_{0}-1\right)
=8$), therefore we do not discuss the results for $N_{\mathrm{even}}$ in
detail here. It is relevant to point out that, in the case of $\gamma =0$
(no background loss), one can again obtain an explicit relation between the
linear gain $\Gamma _{1}$ and the inverse width $\lambda _{r}$, which is
similar to Eq. (\ref{eq:exact1}):
\begin{equation*}
\Gamma _{2}+\frac{B\Gamma _{1}}{E}=\frac{\pm \sinh \lambda _{r}}{1\pm
e^{-2\lambda _{r}(N_{0}-1/2)}}.
\end{equation*}%
Here, as before, the plus and minus signs correspond to $\lambda _{i}=0$ and
$\lambda _{i}=\pi $, respectively.

\begin{figure}[tbph]
\begin{center}
\includegraphics[width = 100mm, keepaspectratio]{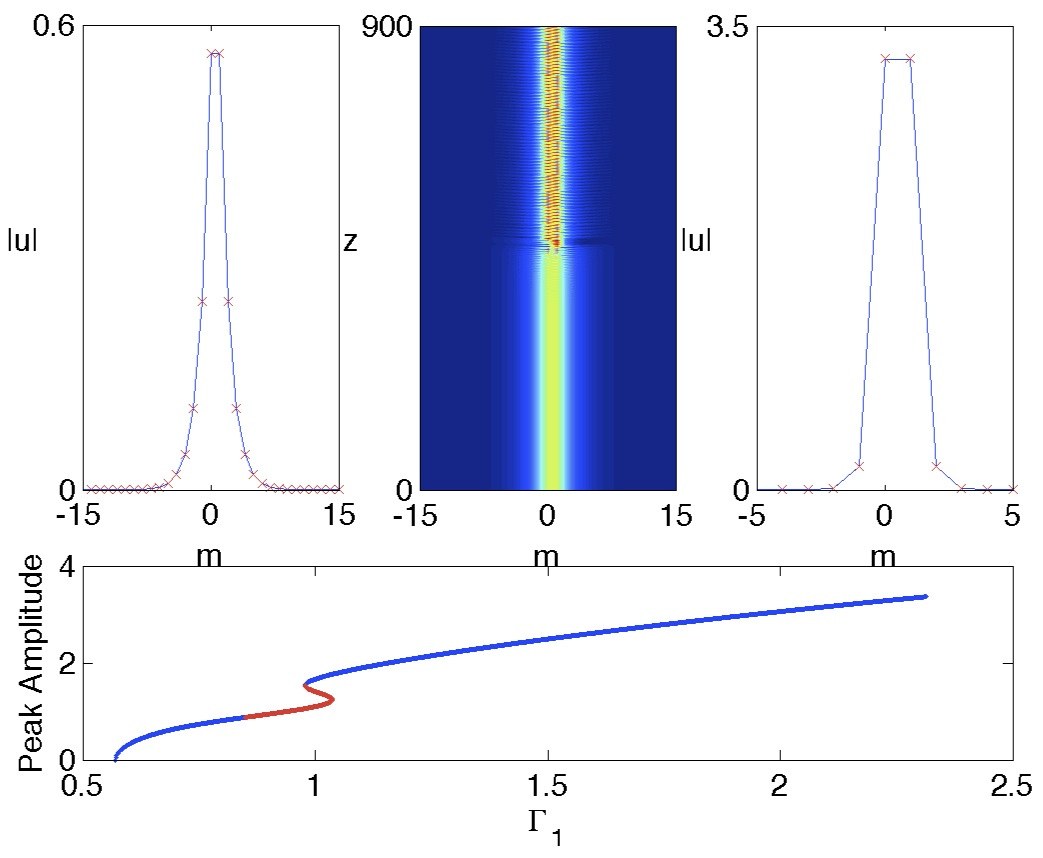}
\end{center}
\caption{(Color online) The top left and right panels: examples of stable
stationary modes supported by two adjacent hot spots (as obtained from both
direct simulations and solution of the reduced system (\protect\ref{eq:sym2}%
)), at $\Gamma _{1}=0.6656$ and $\Gamma _{1}=2.186$, respectively. The top
middle panel: a breather revealed by the direct simulations at $\Gamma
_{1}=0.9099$. Other parameters are $N_{0}=1$, $B=-1$, $E=0.16$, $\Gamma
_{2}=0.8$, and $\protect\gamma =0.5$. The bottom panel displays the
corresponding solution branch, with an unstable (red) segment in the middle.}
\label{fig12}
\end{figure}

A specific particular case is the one when the two HSs are set at adjacent
sites of the lattice, without intermediate sites between them ($N_{\mathrm{%
even}}=0$), which corresponds to $N_{0}=1$ in Eqs. (\ref{N}) and (\ref%
{eq:ansatz2}). Figure~\ref{fig12} shows examples of stationary solutions for
this case. Obviously, their shape does not feature distinct peaks, in
contrast with the profile depicted in Fig.~\ref{fig3}. However, the
stability is similar to that for $N_{0}>1$, stable solutions being found for
small and large values of $\Gamma _{1}$, while breathers are observed at
intermediate values of $\Gamma _{1}$, as a result of the Hopf bifurcation,
see Fig. \ref{fig12}.

\section{Conclusion}

We have introduced an analytically tractable discrete dissipative model, in
which stationary modes are supported by a symmetric pair of embedded HSs
(hot spots), represented by two sites carrying the linear gain,
self-defocusing nonlinearity, and cubic loss. The system can be readily
implemented in photonics and plasmonics, using waveguiding arrays, with the
gain and nonlinearity applied to two cores, which are chosen as HSs.
Symmetric solutions were obtained in an implicit analytical form, separately
for odd and even numbers of intermediate sites between the two HSs, $N_{%
\mathrm{odd}}$ and $N_{\mathrm{even}}$. The latter case includes the
configuration with adjacent HSs ($N_{\mathrm{even}}=0$). The modes with $N_{%
\mathrm{odd}}$ feature both the simple shape, with a minimum at the center,
and the W-shaped one, with an additional lower peak at the central point.
The linear stability of the stationary solutions was investigated \ in the
numerical form, which has revealed both stable and unstable portions of the
solution families. In most cases when the instability is accounted for by
isolated real positive eigenvalues, direct simulations demonstrate that the
unstable mode spontaneously evolves into a stable one, which can be found at
the same parameter values. However, in some cases the evolution transforms
unstable symmetric modes into apparently antisymmetric stable solutions. The
instability represented by a pair of complex conjugate eigenvalues gives
rise to persistent breathers.

The same model may give rise to antisymmetric solutions, detailed results
for which will be reported elsewhere (here, it has been found that some
unstable symmetric modes evolve into stable antisymmetric ones). A
challenging issue, which will be studied separately too, is to search for
asymmetric modes, with unequal amplitudes at the positions of the two
symmetric HSs. Another challenging direction is to extend the analysis for a
similar two-dimensional system.

\section*{Acknowledgment}

A partial financial support for this project has been provided by the
University of Hong Kong Incentive Award Scheme.

\end{document}